# The structure of human olfactory space

Alexei A. Koulakov[1*], Armen G. Enikolopov[1,2], and Dmitry Rinberg[3]

[1]Cold Spring Harbor Laboratory, Cold Spring Harbor, NY 11724
[2]Department of Biological Sciences, Columbia University, New York, NY 10027
[3]HHMI Janelia Farm Research Campus, HHMI, Ashburn, VA 20147
*Correspondence: koulakov@cshl.edu

We analyze the psychophysical responses of human observers to an ensemble of monomolecular odorants. Each odorant is characterized by a set of 146 perceptual descriptors obtained from a database of odor character profiles. Each odorant is therefore represented by a point in highly multidimensional sensory space. In this work we study the arrangement of odorants in this perceptual space. We argue that odorants densely sample a two-dimensional curved surface embedded in the multidimensional sensory space. This surface can account for more than half of the variance of the psychophysical data. We also show that only 12% of experimental variance cannot be explained by curved surfaces of substantially small dimensionality (<10). We suggest that these curved manifolds represent the relevant spaces sampled by the human olfactory system, thereby providing surrogates for olfactory sensory space. For the case of 2D approximation, we relate the two parameters on the curved surface to the physico-chemical parameters of odorant molecules. We show that one of the dimensions is related to eigenvalues of molecules' connectivity matrix, while the other is correlated with measures of molecules' polarity. We discuss the behavioral significance of these findings.

## INTRODUCTION

Our current understanding of many sensory modalities is based on knowledge of the underlying sensory spaces. For example, visual stimuli are well described by their position and the spectral content of the light emitted/scattered by them. The somatosensory system represents positions of stimuli relative to the body surface, which leads to the body-centric somatosensory world. Our understanding of the sense of smell is hindered by the lack of a well-defined perceptual space and knowledge of how this space is related to the properties of odorant molecules[1-2].

It is quite easy to build a generalist olfactory space. Every monomolecular component of a complex odor can be viewed as an individual dimension with the coordinate along this dimension determined by a concentration of the given component. The number of dimensions in this space is equal to the number of monomolecular chemical compounds that can be presented to the olfactory system. Every odorant can be placed uniquely in such a generalist olfactory space. Although the space is complete, it is too large to be relevant to human observers. This is simply because the brain does not have enough olfactory receptors to sample the generalist space in its entirety. The olfactory system is likely to sample a more specialist, lower dimensionality subspace of the general space. The structure of the relevant sensory subspace can be derived from various properties of the olfactory system and is the topic of this and prior studies[3-5].

Here we investigate the structure of olfactory space defined by the responses of human observers. We base our analyses on the Atlas of Odor Character Profiles (AOCP)[6], a database of psychophysical responses of human observers to an array of odorants. In the course of our analyses we discovered that odorants in human olfactory space accumulate near a 2D curved manifold (a curved surface that can be locally approximated by a plane).

The 2D manifold accounted for 51% of the variability in the experimental data. This finding prompted us to seek an approximation to the sensory space in the form of curved continuous surfaces of higher dimension. We show that an approximation of these responses with continuous spaces of sufficiently low dimensionality higher than two could account for 81% of the variability in experimental data. We also find that the intrinsic statistical variability in the data is at least 7%. Thus, only the remaining variance of 12% or less can be attributed to discontinuous features in the sensory space. We argue therefore that a curved continuous manifold of sufficiently low dimension carries most of the information about known features of human olfactory perception.

**RESULTS**

The AOCP database contains information about responses of human observers to 144 monomolecular odorants. Each odorant is characterized by a set of 146 psychophysical descriptors, such as 'fruity', 'floral', 'sickening', 'warm', etc. (see Supplementary Material for complete list of odorants and descriptors ). Thus, each odorant/descriptor pair is characterized by the percentage of observers that recognized the descriptor as applying to odorant. The database can therefore be viewed as a set of 144 points representing individual odorants positioned in a 146-dimensional space of psychophysical descriptors. The resultant cloud of 144 points placed into the multidimensional space of descriptors contains vast information about human perception of monomolecular odorants.

To visualize the multidimensional cluster of odorants we used principal component analysis (PCA). This technique determines directions in the 146D descriptor space that account for the greatest proportion of data variance. These directions [principal components (PCs)] are recognized as the least redundant dimensions and thereby the most informative about the data set[7]. The cluster of 144 odorants projected onto a 3D space specified by the first three principal components is shown in Figure 1A and B. To characterize the amount of information missing from the PCA projection it is customary to present the fraction of variance that is included in the low-dimensional representation (Figure 1E). The 3D PCA projection in Figure 1A includes 52% of the variance in the original data. 48% of the data variance is therefore excluded from 3D representation. The value of PCA in our analysis is in visualizing the correlations present in the data rather than in accounting for these correlations.

Odorants projected to 3D PC space when viewed from certain direction clustered near a C-shaped curve, suggesting that the data points reside close to a 2D surface (Figure 1B). We therefore fitted the set of points by the smooth curved surface shown in Figure 1C and D. The best fit was obtained in 146D space by minimizing the distances from the data points to the nearest points on the surface. To capture the curvature of the surface, it was defined by a second-order polynomial function of 2 parameters: the first PC and a linear combination of the second and the third PCs. The 2D curved surface (manifold) accounted for 94% of the data variance projected to the three principal components (Figure 1A and B) and 56% of the data variance contained in the entire data set containing 146 dimensions.

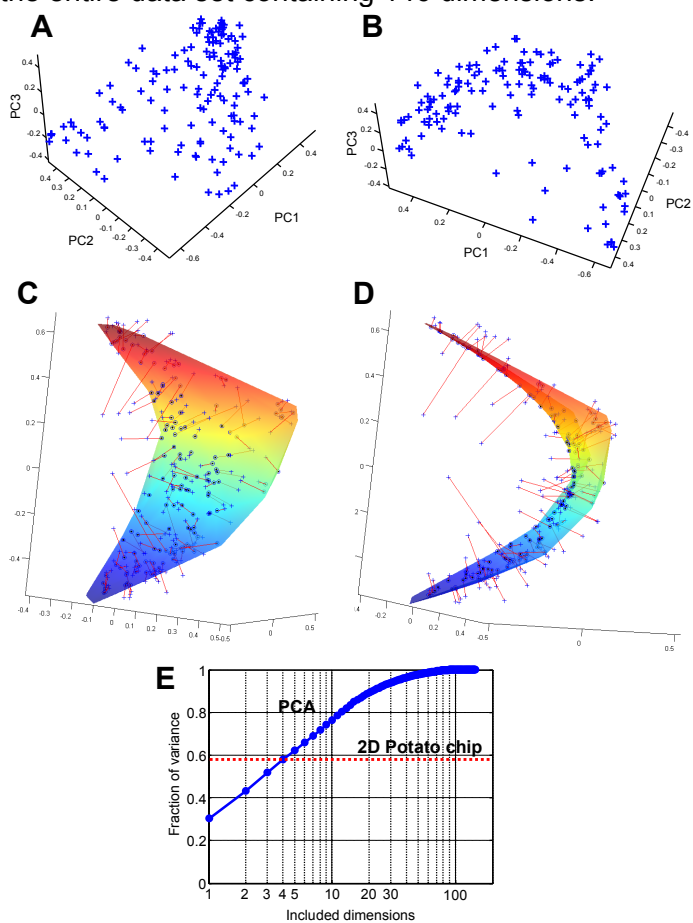

**Figure 1.** Odorants in the PCA space. (A) Each of 144 odorants can be represented as a point in the 146D space of psychophysical descriptors. The odorants are shown by blue crosses placed in the 3D space of principal components. (B) When viewed from certain direction, the odorants clustered near a C-shaped 1D curve, suggesting that in 3D the odorants are distributed close to a 2D curved surface. (C and D) The 2D surface representing the best fit to the data. The odorants (blue crosses) are connected to the nearest points on the surface by the red lines representing the residual errors. The 2D surface minimizes the total squared length of the residuals computed in 146D. The total squared length of residuals can be viewed as remaining variance in the data not accounted for by the projection to the 2D curved manifold. (E) The fraction of included variance as a function of the number of PCA dimensions. The fraction of variance accounted by the 2D curved manifold in (C) and (D) is 56% (red dotted line).

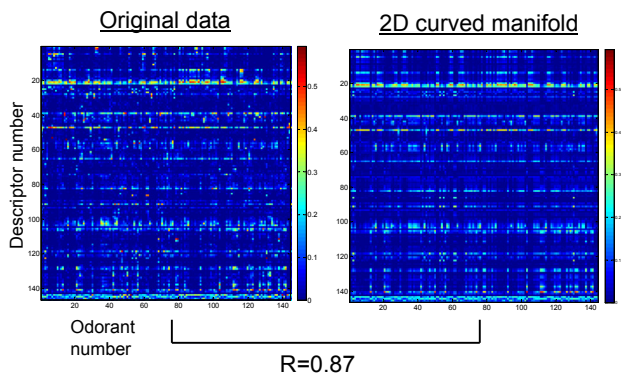

**Figure 2.** Comparison between original psychophysical data and its projection on to 2D curved manifold. Images represent the coordinates (color coded) of 144 odorants in 146D space of descriptors for (A) original data, blue crosses in Figure 1 C&D and (B) its projections to 2D curved manifold, circles in Figure 1 C&D.

How well does a 2D curved manifold in 146D space predict the responses of human observers? To answer this question we compared the original data and its projection to a 2D curved surface (Figure 2). The projections were defined as the nearest points on the 2D surface to a given odorant, as illustrated in Figure 1C and D. The comparison of the two sets of points yielded a correlation coefficient of 87%. Because some correlation is introduced by the average responses to a given descriptor (horizontal bands in Figure 2), we also obtained the correlation coefficient when the averages are excluded from the matrices. This procedure resulted in a correlation coefficient of 75% between the original data and the 2D projection. We conclude that the 2D curved space yields a potent approximation to the psychophysical responses of human observers and therefore forms a reliable surrogate for human olfactory sensory space. We next determined what descriptors contribute to the two parameters on the surface. The first parameter (elevation) is associated with the first PC of the data. As has previously been suggested, this parameter could be correlated with the pleasantness or perceptual valence of odorants[3-4,8]. Consistent with this observation, we find that the psychophysical descriptors that contribute to the first coordinate with large positive/negative coefficients are associated with repulsive/attractive odorant properties (see top and bottom of Figure 3 for the 10 descriptors with the largest positive/negative coefficients, respectively). The second coordinate on the 2D manifold (azimuth) was obtained as a linear combination of the second and the third PCs. The psychophysical descriptors contributing with large coefficients to this coordinate are listed in Figure 3 too (left and right). A possible significance of the second coordinate is discussed below.

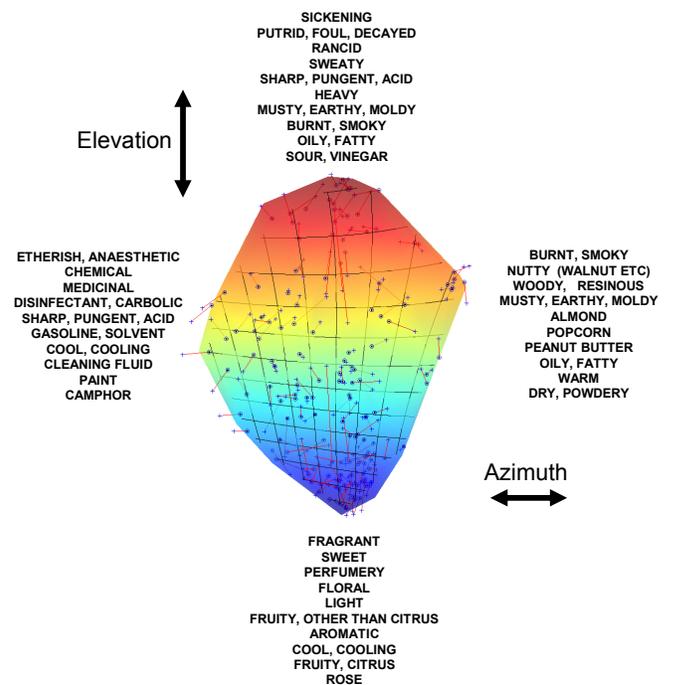

**Figure 3.** Psychophysical descriptors that contribute with large positive/negative coefficients to the coordinates on the 2D surface. The two coordinates are defined as elevation and azimuth as indicated.

Could a curved manifold of dimensionality higher than two characterize human olfactory space more fully? Because we use second order polynomials in our approximation, the number of parameters of the regression is proportional to square of the number of dimensions. To avoid overfitting, we used the jackknife procedure[9] (see Methods for details). In this procedure, a single odorant is removed from the database, an approximation is calculated based on remaining entries in the database, and the result is compared with the odorant that is left out. Our results show that a space of sufficiently small dimensionality ($\leq 10$) can account for a substantial fraction of variance in the experimental data (up to 81%, Figure 4A). The resulting continuous spaces also allow a substantial correlation with experimental data. Thus, when the experimental data are 'projected' to smooth curved manifolds of varying dimensionality by finding the nearest points on the manifold, the correlation between the data and projections can reach 90% or 94% (Figure 4B). These two results are obtained for data sets in which the means of the rows are subtracted/included respectively (Figure 2) and, therefore the data are centered/non-centered with respect to each of the 146 perceptual dimensions.

We found therefore that about 81% of the variance in the dataset is captured by the smooth curved manifolds. We also estimated the errors present in the data due to a finite number of human subjects contributing to the dataset to be about 7%. We conclude that only about 12% of the variance in the experimental data cannot be captured by continuous curved manifolds of dimensionality $\leq 10$. Most (51%) of the experimental variance is reproduced by the 2D curved surface considered above.

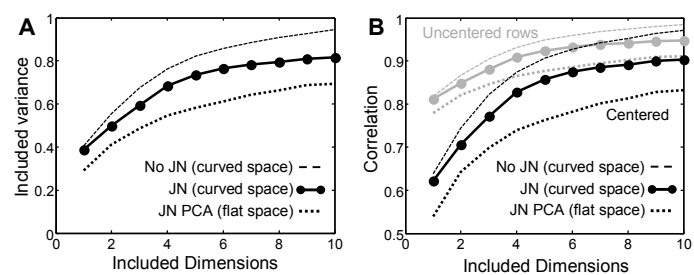

**Figure 4.** Approximation of psychophysical responses with spaces of small dimensionality. To avoid overfitting we applied jackknife (JN) technique. The results for best curved/flat spaces are shown by solid/dotted lines as a function of number of dimensions included. The flat space technique is equivalent to PCA and is shown for comparison. (A) The variance of dataset accounted by the low-dimensional representation. 2D curved manifold accounted for 51% of experimental variance. (B) Pearson correlation as a function of surface dimensionality.

We then attempted to establish the relationship between the two perceptual dimensions (elevation and azimuth) and the physico-chemical properties of odorants. To this end it is necessary to refine the definition of perceptual coordinates on the surface. As seen in Figure 3, the odorants tend to accumulate near the poles of the 2D surface (large positive and negative values of elevation). To remove this singularity we found a non-linear (quadratic) transformation that makes the density of odorants approximately uniform throughout the surface. The new coordinate grid is displayed in Figure 3 on the 2D manifold. The resulting two coordinates on the surface, elevation and azimuth, were then compared to various physico-chemical and structural properties of odorants. 72 physico-chemical properties were obtained from the computer package Molecular Modeling Pro[10]. The structural descriptors included 7 atom counts, 16 pair counts, and 31 triples counts obtained from structural formulas of odorants. The total physico-chemical/structural space included 126 properties for each molecule. We then applied a greedy algorithm developed by Refs. [9,11] to find which properties correlate best with the two perceptual dimensions. The greedy algorithm is an iterative procedure that increases the number of included properties one by one. On each step a new property is added if it results in a maximum increase of Pearson correlation coefficient with a given perceptual dimension. The results of this analysis, physico-chemical properties that yield the best correlation with both azimuth and elevation are presented in Table 1 as a function of the number of included physico-chemical properties (iteration steps).

**Table 1.** Physico-chemical and structural properties of odorant molecules that contribute most strongly to the two perceptual dimensions. CIM is Chemical intuitive molecular index. Order is the number of dimensions included. R is Pearson correlation coefficient.

| | Elevation | | Azimuth | |
|---|---|---|---|---|
| Order | Name | R | Name | R |
| 1 | Burden CIM8 | 0.53 | Water of hydration | 0.33 |
| 2 | C-S pairs | 0.59 | Normal melting point | 0.38 |
| 3 | N count | 0.63 | Debye dipole moment | 0.42 |
| 4 | Molecular width | 0.66 | H-S pairs | 0.44 |
| 5 | H-C-O triples | 0.68 | Dispersion 3D | 0.47 |

The elevation coordinate on the surface is correlated with structural properties of the molecules, such as Burden chemical intuitive molecular indices (CIMs), which represent eigenvalues of the connectivity matrix[10]. These eigenvalues represent simple surrogates for the solution of the quantum-mechanical Hamiltonian equation. We found that all CIMs (1 through 10) are generally well correlated with the elevation coordinate. We also found that simple carbon atom number yields almost the same correlation as CIMs (R=0.50, see Supplementary Material for more detail). For the azimuth coordinate we find that the correlated variables are descriptive of molecules' polarity or hydrophobicity. Thus, four of the five best correlated properties in Table 1 for azimuth depend on molecules' polarity, including melting point temperature. We conclude that the azimuth on the 2D curved manifold is correlated with the hydrophobicity or polarity of odorant molecules.

**DISCUSSION**

In this study we showed that a smooth curved surface of substantially small dimensionality can successfully approximate the responses of human observers to a variety of monomolecular odorants. A 2D curved surface can account for most of the variance in behavioral data. In agreement with previous studies[3-4,8,12], we suggest that one of the dimensions on the 2D surface is the pleasantness or perceptual valence of the odorants. This dimension is correlated with some physico-chemical properties of the molecules, such as the count of the carbon atom count or eigenvalues of the connectivity matrix associated with the structural formula (CIMs[10]). The second perceptual dimension is correlated with the measures of polarity or hydrophobicity, such as water of hydration, normal melting point temperature, etc (Table 1). Because mammalian Class I olfactory receptors (ORs) are related to fish ORs that are expected to bind water-soluble compounds[13], the second dimension may be detected by the difference in responses of the two classes of olfactory receptors: Class I and II. The importance of differentiation between two classes of receptors is highlighted in the mammalian olfactory system by their anatomical segregation[14-15]. The perceptual significance of this second coordinate (dimension) is less straightforward.

An intriguing possibility for the second perceptual coordinate is suggested by studies of cross-modal correlations of smells and sounds. It is reported for example that some Amazonian tribes recognize synesthetic coupling of music and smells[16]. An association between auditory pitch and odorant quality has been proposed for a long time[17-19]. The arrangement of smells as a function of auditory frequency or pitch was shown to be consistent between human subjects[12]. Most importantly, this arrangement was shown to be independent of the pleasantness of odorants[12]. The latter observation suggests an interpretation of the second olfactory dimension (azimuth) as related to the auditory pitch of sounds synesthetically associated with the odorants. This is because this dimension is perpendicular (decorrelated) to pleasantness (elevation) and is the second most significant dimension of olfaction. In agreement with this interpretation, the second dimension emerges from differentiation between groups of psychophysical descriptors that include 'burnt', 'oily', 'fatty' and the group including 'etherish', 'chemical', and 'medicinal'.

The two dimensions of olfaction studied here could mediate two different escape behaviors important for humans. Thus, the pleasantness dimension (elevation) could mediate the escape from products of bacterial decay. Descriptors that are important to this dimension include 'putrid', 'foul', 'decayed', 'rancid', etc. On the other hand, the second dimension (azimuth) could be important for detection of carcinogens in burnt food. These substances known as heterocyclic amines are formed during the cooking of meat, by condensation of creatinine with amino acids[20]. The set of descriptors significant for the second dimension ('burnt', 'smoky') is in agreement with this interpretation.

The low dimensionality of the olfactory space reported here does not eliminate the complexity of olfactory percepts. Indeed, if one adopts a 2D approximation to olfactory space, odor identity depends only on two parameters. But the surface buckles into all 146 dimensions due to its curvature. Thus, although a correlation is present in the data that allows us to reduce the dimensionality of the dataset, olfactory percepts remain complex and varying in all 146 dimensions.

We report here that the human perceptual space of monomolecular odorants can be viewed as continuous, curved, and low-dimensional. Most of the variance in the perceptual data is captured by a 2D curved surface. The two dimensions of the surface can be related to physic-chemical properties of odorant molecules such as an eigenvalue of the odorant molecule connectivity matrix and the polarity of the molecules respectively.

**METHODS**

**Responses to odorants.** Responses to 144 odorants were obtained from Ref. [6] and represented in a set of 146D vectors $\vec{r}_i$ (i=1…144). We used PU set of responses from Ref [6]. PU (percent used) describes the fraction of about 150 observers that thought that a given descriptor applies to an odorant. We verified that our conclusions do not change substantially if other parameters are used instead of PU. We performed PCA on the vectors using SVD procedure. All computations were performed using MATLAB (Mathworks, Inc.). Before applying PCA we normalized response vectors to have unit length in terms of $L_2$ measure. This implies that the vectors resided on a unit sphere in 146D. This reduced somewhat the dimensionality of the dataset to 145D. The normalization step was intended to equalize the odorants in their perceived intensity or concentration. We verified that our conclusions do not change qualitatively if other measures ($L_2$ through $L_9$) are used for normalization. We noticed some deterioration of the fits beyond this range.

**Approximating odorant response with curved spaces.** Each odorant vector $\vec{r}_i$ was approximated with the 'projected' vector $\vec{p}_i$. Here index $i$ enumerates the odorants while each vector contains 146 components corresponding to psychophysical descriptors. The projected vectors were sought in the form

$$\vec{p}_i = \vec{A} + \sum_{\alpha=1}^{D} \vec{B}_\alpha x_{\alpha i} + \sum_{\alpha=1}^{D}\sum_{\beta=1}^{D} \vec{C}_{\alpha\beta} x_{\alpha i} x_{\beta i} . \quad (1)$$

Here $\vec{A}$, $\vec{B}_\alpha$, and $\vec{C}_{\alpha\beta}$ are odorant-independent parameters of the surface. Parameters $\vec{C}_{\alpha\beta}$ allowed the surface to be curved. Parameters $x_{\alpha i}$ define positions of odorants on the surface. $D$ is the number of parameters per odorant which is the dimensionality of the surface. The manifold defined by this equation is $D$-dimensional. In figure 2 we used $D=2$, while in Figure 4 the dimensionality was varied. To find $\vec{A}$, $\vec{B}_\alpha$, $\vec{C}_{\alpha\beta}$, and $x_{\alpha i}$ we minimized $\sum_i \|\vec{r}_i - \vec{p}_i\|^2$ using conjugate gradient algorithm. The set of parameters $x_{\alpha i}$ was determined therefore as the nearest points on the curved manifold. The nearest points define 'projections' onto the curved manifold.

**Jackknife procedure.** Approximating human psychophysical responses with higher dimensional curved manifolds is confounded by a dramatic increase in the number of parameters of fit. Because the number of parameters increases as a second power of the number of dimensions in our quadratic regression, for a moderately low-dimensional manifold we find that we can perfectly fit all of the experimental data (Figure 4A, dashed line). To avoid this overfitting problem we employed the jackknife technique, in which we remove a single odorant from the perceptual database, obtain a high-dimensional fit for the responses to the remaining compounds, and calculate the distance between the fitted manifold and the removed odorant. By applying this procedure for all odorants in the database sequentially we evaluated a variance of the approximation with curved manifolds. The variance does not vanish for spaces of high dimensionality due to overfitting (Figure 4A, solid line).

**The natural system of coordinates of the 2D surface** was used to equilibrate the density of odorants (grid in Figure 3). The odorants were projected to 2D PCA space and the Delaunay triangulation was calculated. The edges of triangulation were replaced with elastic strings of unit equilibrium length and a coordinate transformation was found that minimizes the elastic energy of the strings. The coordinate transformation was constrained to the form used above [equation (1)] with the mapping of 2D to 2D space. The results are shown in the Supplementary Material.

**Estimating the variability due to a finite number of observers.** The psychophysical variable used here (percent used, PU) is convenient for estimating the experimental variability. We resampled the data for every entry in the database independently using 149 observers as specified in Ref. [6]. We estimated the variance of the resulting ensemble to be equal to 7% of the experimental variance present in Ref. [6].

**Physico-chemical parameters.** The values of 72 parameters were calculated using the program Molecular Modeling Pro™ (ChemSW, Failfield, CA). We verified that the use of 1999 parameters generated by E-Dragon (VCCLAB.org) did not improve the result suggesting a redundancy in the data. We used logarithms of all the parameter values normalized to a unit variance and zero mean for each parameter across odorants.

*SUPPLEMENTARY METERIAL*

**The Structure of Human Olfactory Space**

by Alexei A. Koulakov, Armen G. Enikolopov, and Dmitry Rinberg

**1. Odorants included in the analysis**

The following odorants were used from the Atlas of Odor Character profiles.

| | | |
|---|---|---|
| 1 | 698-10-2 | Abhexone |
| 2 | 98-86-2 | Acetophenone |
| 3 | 1122-62-9 | Acetyl Pyridine: ortho-Acetyl Pyridine |
| 4 | 141-13-9 | Adoxal |
| 5 | 77-83-8 | Aldehyde C-16 (So-Called)→ Lower Concentration |
| 6 | 77-83-8 | Aldehyde C-16 (So-Called)→ Higher Concentration |
| 7 | 104-61-0 | Aldehyde C-18 (So-Called) |
| 8 | 123-68-2 | Allyl Caproate |
| 9 | 123-82-2 | Amyl Acetate: iso-Amyl Acetate |
| 10 | 540-18-1 | Amyl Butyrate |
| 11 | 60763-41-9 | → Amyl Cinnamic Aldehyde Diethyl Acetal |
| 12 | 102-19-2 | → Amyl Phenyl Acetate |
| 13 | 2173-56-0 | → Amyl Valerate |
| 14 | 29597-36-2 | → Andrane |
| 15 | 104-46-1 | → Anethole |
| 16 | 100-66-3 | → Anisole |
| 17 | 89-43-0 | → Auralva |
| 18 | 100-52-7 | → Benzaldehyde |
| 19 | 119-84-8 | → Benzo Dihydro Pyrone |
| 20 | 5655-61-8 | → Bornyl Acetate: iso-Bornyl Acetate |
| 21 | 107-92-6 | → Butanoic Acid |
| 22 | 71-38-3 | → Butanol: 1-Butanol |
| 23 | 544-40-1 | → Butyl Sulfide |
| 24 | 67634-06-4 | → Butyl Quinoline: iso-Butyl Quinoline |
| 25 | 78-22-2 | → Camphor: dl-Camphor |
| 26 | 99-49-0 | → Carvone: l-Carvone |
| 27 | 87-44-5 | → Caryophyllene (beta and gamma Isomers) |
| 28 | 33704-61-9 | → Cashmeran |
| 29 | 17369-59-4 | → Celeriax |
| 30 | 89-68-9 | → Chlorothymol |
| 31 | 104-55-2 | → Cinnamic Aldehyde |
| 32 | 141-27-5 | → Citral |
| 33 | 5585-39-7 | → Citralva |
| 34 | 91-64-5 | → Coumarin |
| 35 | 108-39-4 | → Cresol: m-Cresol |
| 36 | 106-44-5 | → Cresol: p-Cresol |
| 37 | 140-39-6 | → Cresyl Acetate: p-Cresyl Acetate |
| 38 | 103-93-5 | → Cresyl Butyrate: p-Cresyl-iso-Butyrate |
| 39 | 104-93-8 | → Cresyl Methyl Ether: p-Cresyl Methyl Ether |
| 40 | 122-03-2 | → Cuminic Aldehyde |
| 41 | 1423-46-7 | → Cyclocitral: iso-Cyclocitral |
| 42 | 55704-78-4 | → Cyclodithalfarol |
| 43 | 765-87-7 | → Cyclohexanedione: 1,2-Cyclohexanedione |
| 44 | 108-93-0 | → Cyclohexanol |
| 45 | 80-71-7 | → Cyclotene |
| 46 | 67634-23-5 | → Cyclotropal |
| 47 | 25152-84-5 | → Decadienal: 2,4-trans-trans-Decadienal |
| 48 | 91-17-8 | → Decahydro Naphthalene |
| 49 | 111-92-2 | → Dibutyl Amine |
| 50 | 352-93-2 | → Diethyl Sulfide |
| 51 | 10094-34-5 | → Dimethyl Benzyl Carbinyl Butyrate |
| 52 | 103-05-9 | → Dimethyl Phenyl Ethyl Carbinol |
| 53 | 5910-89-4 | → Dimethyl Pyrazine: 2,3-Dimethyl Pyrazine |
| 54 | 123-32-0 | → Dimethyl Pyrazine:  2,5-Dimethyl Pyrazine |
| 55 | 625-84-3 | → Dimethyl Pyrrole: 2,5-Dimethyl Pyrrole |
| 56 | 3658-80-8 | → Dimethyl Trisulfide |
| 57 | 4747-07-3 | → Diola |
| 58 | 101-84-8 | → Diphenyl Oxide |
| 59 | 105-54-4 | → Ethyl Butyrate |
| 60 | 105-37-3 | → Ethyl Propionate |
| 61 | 13925-00-3 | →  2-Ethyl Pyrazine (Lower Concentration) |
| 62 | 13925-00-3 | → 2-Ethyl Pyrazine (Higher Concentration) |
| 63 | 470-82-6 | → Eucalyptol |
| 64 | 97-53-0 | → Eugenol |
| 65 | 67634-15-5 | → Floralozone |
| 66 | 6413-10-1 | → Fructone |
| 67 | 98-01-1 | → Furfural |
| 68 | 98-02-2 | → Furfuryl Mercaptan |
| 69 | 88683-93-6 | → Grisalva |
| 70 | 90-05-1 | → Guaiacol |
| 71 | 111-71-7 |  → Heptanal |
| 72 | 111-70-6 | → Heptanol: 1-Heptanol |
| 73 | 68-25-1 | → Hexanal |
| 74 | 142-62-1 |    Hexanoic acid |
| 75 | 111-27-3 | → Hexanol: 1-Hexanol |
| 76 | 623-37-0 | → Hexanol: 3-Hexanol |
| 77 | 6728-26-3 | → Hexenal: trans-1-Hexenal |
| 78 | 111-26-2 | → Hexyl Amine (Lower Concentration) |
| 79 | 111-26-2 | → Hexyl Amine (Higher Concentration) |
| 80 | 101-86-0 | → Hexyl Cinnamic Aldehyde |
| 81 | 90-87-9 |  → Hydratropic Aldehyde Dimethyl Acetal |
| 82 | 107-75-5 | → Hydroxy Citronellal |
| 83 | 120-72-9 | → Indole |
| 84 | 67801-36-9 | → Indolene |
| 85 | 75-47-8 | → Iodoform |
| 86 | 14901-07-6 | → Ionone: beta-Ionone (Lower Concentration) |
| 87 | 14901-07-6 | → Ionone: beta-Ionone (Higher Concentration) |
| 88 | 79-69-6 | → Irone: alpha-Irone |
| 89 | 126-91-0 | → Linalool |
| 90 | 138-86-3 | → Limonene: d-Limonene |
| 91 | 31906-04-4 | → Lyral |
| 92 | 67258-87-1 | → Maritima |
| 93 | 106-72-9 | → Melonal |
| 94 | 2216-51-5 | → Menthol: l-Menthol |
| 95 | 93-04-9 | → Methoxy-Naphthalene: 2-Methoxy Naphthalene |
| 96 | 134-20-3 | → Methyl Anthranilate |
| 97 | 462-95-3 | → Methyl Acetaldehyde Dimethyl Acetal |
| 98 | 1334-76-5 | → Methyl Furoate |
| 99 | 2271-428 | → Methyl-iso-Borneol: 2-Methyl-iso-Borneol |
| 100 | 491-35-0 | 0→ Methyl Quinoline: para-Methyl Quinoline |
| 101 | 2459-09-8 | 1 → Methyl iso-Nicotinate |
| 102 | 119-36-8 | 2→ Methyl Salicylate |
| 103 | 2432-51-1 | 3→ Methyl Thiobutyrate |
| 104 | 1222-05-5 | 4→ Musk Galaxolide |
| 105 | 1508-02-1 | 5→ Musk Tonalid |
| 106 | 37677-14-8 | 6→ Myracaldehyde |
| 107 | 143-13-5 | 7→ Nonyl Acetate |
| 108 | 4674-50-4 | 8→ Nootkatone |
| 109 | 111-87-5 | 9→ Octanol: 1-Octanol |
| 110 | 3391-86-4 | 0→ Octenol: 1-Octen-3-OL |
| 111 | 109-52-4 | 1 →  Pentanoic Acid |
| 112 | 591-80-0 | 2→ Pentenoic Acid: 4-Pentenoic Acid |
| 113 | 103-82-2 | 3→ Phenyl Acetic Acid |
| 114 | 536-74-3 | 4→ Phenyl Acetylene |
| 115 | 60-12-8 | 5→ Phenyl Ethanol (Lower Concentration) |
| 116 | 60-12-8 | 6→ Phenyl Ethanol (Higher Concentration) |
| 117 | 78-59-1 | 7→ Phorone: iso-Phorone |
| 118 | 80-56-8 | 8→ Pinene: alpha-Pinene |
| 119 | 105-66-8 | 9→ Propyl Butyrate |
| 120 | 135-79-5 | 0→ Propyl Quinoline: iso-Propyl Quinoline |
| 121 | 111-47-7 | 1 → Propyl Sulfide |
| 122 | 110-86-1 | 2→ Pyridine |
| 123 | 94-59-7 | 3→ Safrole |
| 124 | 69460-08-8 | 4→ Sandiff |
| 125 | 115-71-9 | 5→ Santalol |
| 126 | 83-34-1 | 6→ Skatole |
| 127 | 10482-56-1 | 7→ Terpineol, mostly alpha-Terpineol |
| 128 | 110-01-0 | 8→ Tetrahydro Thiophene |

| # | CAS | Name |
|---|---|---|
| 129 | 91-61-2 | Tetraquinone |
| 130 | 36267-71-7 | Thienopyrimidine |
| 131 | 123-93-3 | Thioglycolic Acid |
| 132 | 110-02-1 | Thiophene |
| 133 | 89-83-8 | Thymol |
| 134 | 529-20-4 | Tolualdehyde: ortho-Tolualdehyde |
| 135 | 108-88-3 | Toluene (Lower Concentration) |
| 136 | 108-88-3 | Toluene (Higher Concentration) |
| 137 | 75-50-3 | Trimethyl Amine |
| 138 | 104-67-6 | Undecalactone: gamma-Undecalactone |
| 139 | 112-38-9 | Undecylenic Acid |
| 140 | 590-86-3 | Valeraldehyde: iso-Valeraldehyde |
| 141 | 503-74-2 | Valeric Acid: iso-Valeric Acid |
| 142 | 108-29-2 | Valerolactone: gamma-Valerolactone |
| 143 | 121-33-5 | Vanillin |
| 144 | 122-48-5 | Zingerone |

## 2. Perceptual descriptors

| | |
|---|---|
| 1 | FRUITY, CITRUS |
| 2 | LEMON |
| 3 | GRAPEFRUIT |
| 4 | ORANGE |
| 5 | FRUITY, OTHER THAN CITRUS |
| 6 | PINEAPPLE |
| 7 | GRAPE JUICE |
| 8 | STRAWBERRY |
| 9 | APPLE (FRUIT) |
| 10 | PEAR |
| 11 | CANTALOUPE, HONEY DEW MELON |
| 12 | PEACH (FRUIT) |
| 13 | BANANA |
| 14 | FLORAL |
| 15 | ROSE |
| 16 | VIOLETS |
| 17 | LAVENDER |
| 18 | COLOGNE |
| 19 | MUSK |
| 20 | PERFUMERY |
| 21 | FRAGRANT |
| 22 | AROMATIC |
| 23 | HONEY |
| 24 | CHERRY (BERRY) |
| 25 | ALMOND |
| 26 | NAIL POLISH REMOVER |
| 27 | NUTTY (WALNUT ETC) |
| 28 | SPICY |
| 29 | CLOVE |
| 30 | CINNAMON |
| 31 | LAUREL LEAVES |
| 32 | TEA LEAVES |
| 33 | SEASONING (FOR MEAT) |
| 34 | BLACK PEPPER |
| 35 | GREEN PEPPER |
| 36 | DILL |
| 37 | CARAWAY |
| 38 | OAK WOOD, COGNAC |
| 39 | WOODY, RESINOUS |
| 40 | CEDARWOOD |
| 41 | MOTHBALLS |
| 42 | MINTY, PEPPERMINT |
| 43 | CAMPHOR |
| 44 | EUCALIPTUS |
| 45 | CHOCOLATE |
| 46 | VANILLA |
| 47 | SWEET |
| 48 | MAPLE SYRUP |
| 49 | CARAMEL |
| 50 | MALTY |
| 51 | RAISINS |
| 52 | MOLASSES |
| 53 | COCONUT |
| 54 | ANISE (LICORICE) |
| 55 | ALCOHOLIC |
| 56 | ETHERISH, ANAESTHETIC |
| 57 | CLEANING FLUID |
| 58 | GASOLINE, SOLVENT |
| 59 | TURPENTINE (PINE OIL) |
| 60 | GERANIUM LEAVES |
| 61 | CELERY |
| 62 | FRESH GREEN VEGETABLES |
| 63 | CRUSHED WEEDS |
| 64 | CRUSHED GRASS |
| 65 | HERBAL, GREEN, CUT GRASS |
| 66 | RAW CUCUMBER |
| 67 | HAY |
| 68 | GRAINY (AS GRAIN) |
| 69 | YEASTY |
| 70 | BAKERY (FRESH BREAD) |
| 71 | SOUR MILK |
| 72 | FERMENTED (ROTTEN) FRUIT |
| 73 | BEERY |
| 74 | SOAPY |
| 75 | LEATHER |
| 76 | CARDBOARD |
| 77 | ROPE |
| 78 | WET PAPER |
| 79 | WET WOOL, WET DOG |
| 80 | DIRTY LINEN |
| 81 | STALE |
| 82 | MUSTY, EARTHY, MOLDY |
| 83 | RAW POTATO |
| 84 | MOUSE |
| 85 | MUSHROOM |
| 86 | PEANUT BUTTER |
| 87 | BEANY |
| 88 | EGGY (FRESH EGGS) |
| 89 | BARK, BIRCH BARK |
| 90 | CORK |

91    BURNT, SMOKY
92    FRESH TOBACCO SMOKE
93    INCENSE
94    COFFEE
95    STALE TOBACCO SMOKE
96    BURNT PAPER
97    BURNT MILK
98    BURNT RUBBER
99    TAR
100    CREOSOTE
101    DISINFECTANT, CARBOLIC
102    MEDICINAL
103    CHEMICAL
104    BITTER
105    SHARP, PUNGENT, ACID
106    SOUR, VINEGAR
107    SAUERKRAUT
108    AMMONIA
109    URINE
110    CAT URINE
111    FISHY
112    KIPPERY (SMOKED FISH)
113    SEMINAL, SPERM
114    NEW RUBBER
115    SOOTY
116    BURNT CANDLE
117    KEROSENE
118    OILY, FATTY
119    BUTTERY, FRESH BUTTER
120    PAINT
121    VARNISH
122    POPCORN
123    FRIED CHICKEN
124    MEATY (COOKED, GOOD)
125    SOUPY
126    COOKED VEGETABLES
127    RANCID
128    SWEATY
129    CHEESY
130    HOUSEHOLD GAS
131    SULFIDIC
132    GARLIC, ONION
133    METALLIC
134    BLOOD, RAW MEAT
135    ANIMAL
136    SEWER

137    PUTRID, FOUL, DECAYED
138    FECAL (LIKE MANURE)
139    CADAVEROUS (DEAD ANIMAL)
140    SICKENING
141    DRY, POWDERY
142    CHALKY
143    LIGHT
144    HEAVY
145    COOL, COOLING
146    WARM

**List of physico-Chemical parameters used**

1. C
2. H
3. O
4. N
5. S
6. I
7. L
8. molecular_weight
9. molecular_volume
10. molecular_length
11. molecular_width
12. molecular_depth
13. density
14. surface_area
15. Log_Kow_fragments
16. HLB
17. solubility_parameter
18. dispersion_3D
19. polarity_3D
20. hydrogen_bond_3D
21. hydrogen_bond_acceptor
22. hydrogen_bond_donor
23. dipole_moment_debye
24. hydrophilic_surface_area
25. water_of_hydration
26. boiling_point_C
27. vapor_pressure_torr
28. MR
29. parachor
30. connectivity_0
31. connectivity_1
32. connectivity_2
33. connectivity_3
34. connectivity_4
35. valence_0
36. valence_1
37. valence_2
38. valence_3
39. valence_4
40. kappa_2
41. log_water_solubility
42. Log_P__atom_based
43. Z_chain_length
44. glass_transition_temperature

| # | Property |
|---|---|
| 45 | melt_transition_temperature |
| 46 | water_content_30_RH |
| 47 | water_content_50_RH |
| 48 | water_content_70_RH |
| 49 | water_content_90_RH |
| 50 | water_content_100_RH |
| 51 | molar_volume |
| 52 | Surface_tension |
| 53 | Viscosity_cp_at_25C |
| 54 | Surface_tension_in_water |
| 55 | Critical_Temperature_K |
| 56 | Critical_Pressure_bar |
| 57 | Normal_Boiling_Point_K |
| 58 | Normal_Freezing_Point_K |
| 59 | Enthalpy_of_formation |
| 60 | Gibbs_energy_of_formation |
| 61 | enthalpy_of_vaporization |
| 62 | enthalpy_of_fusion |
| 63 | liquid_viscosity |
| 64 | heat_capacity_25C |
| 65 | Effective_number_of_torsional_bonds |
| 66 | hydrogen_bond_number |
| 67 | Entropy_of_boiling_JKmol |
| 68 | Heat_capacity_change_on_boiling_JKmol |
| 69 | CIM_1 |
| 70 | CIM_2 |
| 71 | CIM_3 |
| 72 | CIM_4 |
| 73 | CIM_5 |
| 74 | CIM_6 |
| 75 | CIM_7 |
| 76 | CIM_8 |
| 77 | CIM_9 |
| 78 | CIM_10 |
| 79 | Polar_surface_area |
| 80 | C1C |
| 81 | C1H |
| 82 | C1O |
| 83 | C1N |
| 84 | C1S |
| 85 | C1I |
| 86 | C1L |
| 87 | H1O |
| 88 | H1N |
| 89 | H1S |
| 90 | S1S |
| 91 | C2C |
| 92 | C2O |
| 93 | C2N |
| 94 | C3C |
| 95 | C3N |
| 96 | C1C1C |
| 97 | C2C1C |
| 98 | C1C1H |
| 99 | C2C1H |
| 100 | C3C1H |
| 101 | C1C1O |
| 102 | C1C2O |
| 103 | C2C1O |
| 104 | C1C1N |
| 105 | C1C2N |
| 106 | C1C3N |
| 107 | C2C1N |
| 108 | C1C1S |
| 109 | C2C1S |
| 110 | C2C1L |
| 111 | C1O1C |
| 112 | C1O1H |
| 113 | C1N1C |
| 114 | C2N1C |
| 115 | C1N1H |
| 116 | C1S1C |
| 117 | C1S1S |
| 118 | H1C1O |
| 119 | H1C2O |
| 120 | H1C1N |
| 121 | H1C2N |
| 122 | H1C1S |
| 123 | O1C1O |
| 124 | O2C1O |
| 125 | O1C1S |
| 126 | S1S1S |

The data above contains atom counts per molecule (C through L (chlorine)), number of pairs per molecule (C1C or C-C trough C3N or C≡N), and numbers of triples per molecule (C1C1C for C-C-C through S-S-S).

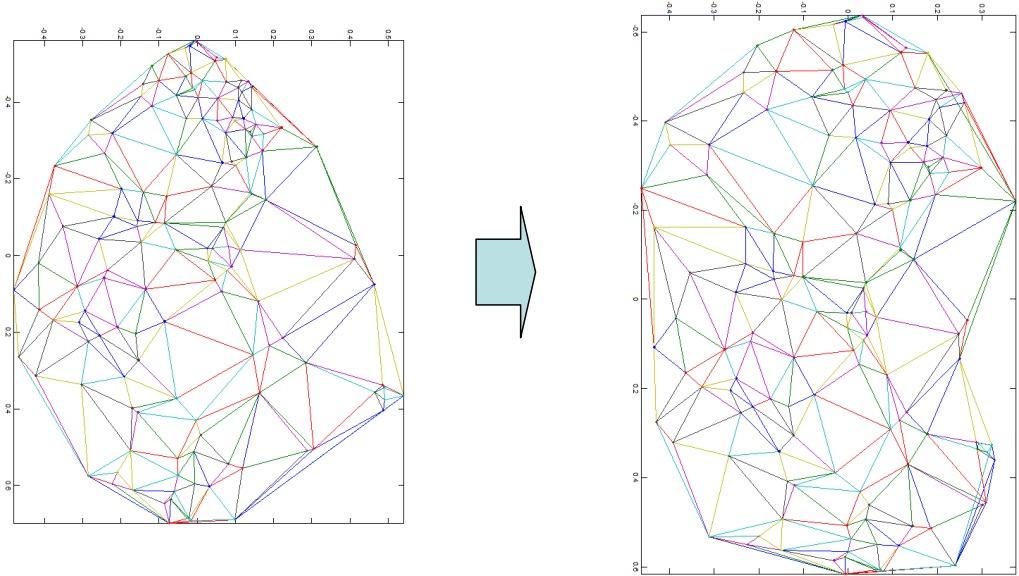

**Supplementary Figure 1.** Equilibrating the density of the odorants in two dimensions. Left: the original set of odorants projected onto a flat 2D space and Delaunay triangulated. Right: the same set of odorants after relaxing the elastic energy of edges that are assumed to be springs with unit equilibrium length and the same elastic coefficient. The transformation (arrow) was constrained to be of the second order as in equation (1) of the main text. The final two coordinates were used to correlate with the structural and physico-chemical parameters (Table 1, Supplementary figure 2).

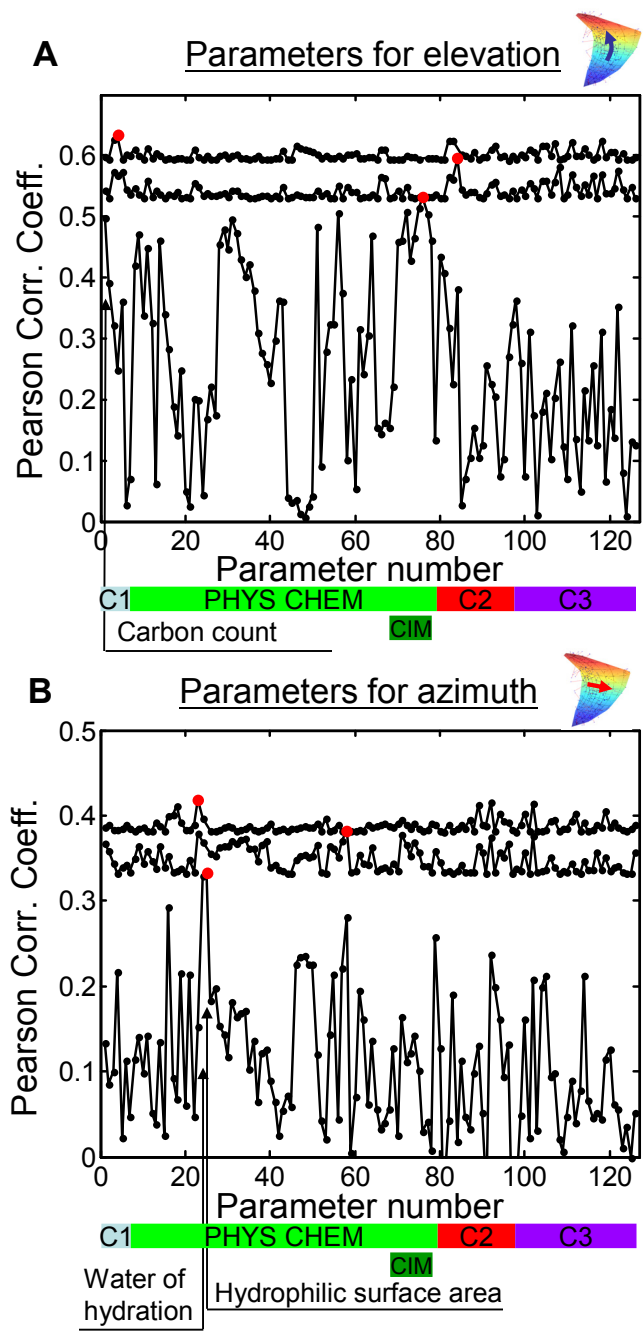

**Supplementary Figure 2.** The results of greedy algorithm for elevation (A) and azimuth (B) variables on the 2D fit to psychophysical data. Pearson correlation coefficient is shown as a function of the number of physico-chemical/structural parameter (see above). Three iterations are shown for each parameter by three lines with dots. The parameters yielding maximal correlation on each iteration are shown by the red dots. Some parameters are highlighted, such as Carbon count (R=0.50), hydrophilic surface area (R=0.33), and water of hydration (R=0.33). Horizontal axis also contains marking indicating the corresponding block of parameters included: element counts (C1), Molecular modeling Pro physico-chemical parameters (PHYS CHEM), pairs counts (C2), and triples counts (C3). CIM is the block of ten Burden chemical intuitive indexes.